%% Testing the EPR Locality using B -Mesons
%% T. Ichikawa, S. Tamura and I. Tsutsui
%% May 2008: revised July 2008
%% Plain TeX

%%% First some fonts %%%%%%%%%%%%%%%%%%%%%%%%%%%%%%%%%
\font\bigbold=cmbx12
\font\bigms=cmmi12
\font\ninerm=cmr9      \font\eightrm=cmr8    \font\sixrm=cmr6
\font\fiverm=cmr5
\font\ninebf=cmbx9     \font\eightbf=cmbx8   \font\sixbf=cmbx6
\font\fivebf=cmbx5
\font\ninei=cmmi9      \skewchar\ninei='177  \font\eighti=cmmi8
\skewchar\eighti='177  \font\sixi=cmmi6      \skewchar\sixi='177
\font\fivei=cmmi5
\font\ninesy=cmsy9     \skewchar\ninesy='60  \font\eightsy=cmsy8
\skewchar\eightsy='60  \font\sixsy=cmsy6     \skewchar\sixsy='60
\font\fivesy=cmsy5     \font\nineit=cmti9    \font\eightit=cmti8
\font\ninesl=cmsl9     \font\eightsl=cmsl8

\font\ninett=cmtt9     \font\eighttt=cmtt8
\font\tenfrak=eufm10   \font\ninefrak=eufm9  \font\eightfrak=eufm8
\font\sevenfrak=eufm7  \font\fivefrak=eufm5
\font\tenbb=msbm10     \font\ninebb=msbm9    \font\eightbb=msbm8
\font\sevenbb=msbm7    \font\fivebb=msbm5
\font\tenssf=cmss10    \font\ninessf=cmss9   \font\eightssf=cmss8
\font\tensmc=cmcsc10

%%% Some Families %%%%%%%%%%%%%%%%%%%%%%%%%%%%%%%%%%%%
\newfam\bbfam   \textfont\bbfam=\tenbb \scriptfont\bbfam=\sevenbb
\scriptscriptfont\bbfam=\fivebb  \def\Bbb{\fam\bbfam}
\newfam\frakfam  \textfont\frakfam=\tenfrak \scriptfont\frakfam=%
\sevenfrak \scriptscriptfont\frakfam=\fivefrak  \def\frak{\fam\frakfam}
\newfam\ssffam  \textfont\ssffam=\tenssf \scriptfont\ssffam=\ninessf
\scriptscriptfont\ssffam=\eightssf  
\def\smc{\tensmc}

%%% Definition of 8 point %%%%%%%%%%%%%%%%%%%%%%%%%%%%
\def\eightpoint{\textfont0=\eightrm \scriptfont0=\sixrm
\scriptscriptfont0=\fiverm  \def\rm{\fam0\eightrm}%
\textfont1=\eighti \scriptfont1=\sixi \scriptscriptfont1=\fivei
\def\oldstyle{\fam1\eighti}\textfont2=\eightsy
\scriptfont2=\sixsy \scriptscriptfont2=\fivesy
\textfont\itfam=\eightit         \def\it{\fam\itfam\eightit}%
\textfont\slfam=\eightsl         \def\sl{\fam\slfam\eightsl}%
\textfont\ttfam=\eighttt         \def\tt{\fam\ttfam\eighttt}%
\textfont\frakfam=\eightfrak     \def\frak{\fam\frakfam\eightfrak}%
\textfont\bbfam=\eightbb         \def\Bbb{\fam\bbfam\eightbb}%
\textfont\bffam=\eightbf         \scriptfont\bffam=\sixbf
\scriptscriptfont\bffam=\fivebf  \def\bf{\fam\bffam\eightbf}%
\abovedisplayskip=9pt plus 2pt minus 6pt   \belowdisplayskip=%
\abovedisplayskip  \abovedisplayshortskip=0pt plus 2pt
\belowdisplayshortskip=5pt plus2pt minus 3pt  \smallskipamount=%
2pt plus 1pt minus 1pt  \medskipamount=4pt plus 2pt minus 2pt
\bigskipamount=9pt plus4pt minus 4pt  \setbox\strutbox=%
\hbox{\vrule height 7pt depth 2pt width 0pt}%
\normalbaselineskip=9pt \normalbaselines \rm}

%%% Definition of 9 point %%%%%%%%%%%%%%%%%%%%%%%%%%%%
\def\ninepoint{\textfont0=\ninerm \scriptfont0=\sixrm
\scriptscriptfont0=\fiverm  \def\rm{\fam0\ninerm}\textfont1=\ninei
\scriptfont1=\sixi \scriptscriptfont1=\fivei \def\oldstyle%
{\fam1\ninei}\textfont2=\ninesy \scriptfont2=\sixsy
\scriptscriptfont2=\fivesy
\textfont\itfam=\nineit          \def\it{\fam\itfam\nineit}%
\textfont\slfam=\ninesl          \def\sl{\fam\slfam\ninesl}%
\textfont\ttfam=\ninett          \def\tt{\fam\ttfam\ninett}%
\textfont\frakfam=\ninefrak      \def\frak{\fam\frakfam\ninefrak}%
\textfont\bbfam=\ninebb          \def\Bbb{\fam\bbfam\ninebb}%
\textfont\bffam=\ninebf          \scriptfont\bffam=\sixbf
\scriptscriptfont\bffam=\fivebf  \def\bf{\fam\bffam\ninebf}%
\abovedisplayskip=10pt plus 2pt minus 6pt \belowdisplayskip=%
\abovedisplayskip  \abovedisplayshortskip=0pt plus 2pt
\belowdisplayshortskip=5pt plus2pt minus 3pt  \smallskipamount=%
2pt plus 1pt minus 1pt  \medskipamount=4pt plus 2pt minus 2pt
\bigskipamount=10pt plus4pt minus 4pt  \setbox\strutbox=%
\hbox{\vrule height 7pt depth 2pt width 0pt}%
\normalbaselineskip=10pt \normalbaselines \rm}

%%% bold font for math vectors %%%%%%%%%%%%%%%%%%%%%%%%%%%%%%%
\font\tensvec=cmmib10 
\font\sevensvec=cmmib10 at 7pt
\font\fivesvec=cmmib10 at 5 pt
\def\svec#1{{\mathchoice%
  {\hbox{$\displaystyle\textfont1=\tensvec%
                       \scriptfont1=\sevensvec%
                       \scriptscriptfont1=\fivesvec#1$}}%
  {\hbox{$\textstyle\textfont1=\tensvec%
                    \scriptfont1=\sevensvec%
                    \scriptscriptfont1=\fivesvec#1$}}%
  {\hbox{$\scriptstyle\scriptfont1=\sevensvec%
                      \scriptscriptfont1=\fivesvec#1$}}%
  {\hbox{$\scriptscriptstyle\scriptscriptfont1=\fivesvec#1$}}%
}}

%%% Macro to generate the equation #'s automatically.
%%% To use start each new section (eg 3) with the commands
%%% \secno=3 \meqno=1 :this will start the equations with (3.1)
%%% Then in place of \eqno(3.1) the type \eqn\descriptivename . To refer
%%% back to the equation simply the type (\descritivename)
%%% For the appendixset \secno=0, \appno=1\meqno=1 etc
%%%
\global\newcount\secno \global\secno=0 \global\newcount\meqno
\global\meqno=1 \global\newcount\appno \global\appno=0
\newwrite\eqmac \def\romappno{\ifcase\appno\or A\or B\or C\or D\or
E\or F\or G\or H\or I\or J\or K\or L\or M\or N\or O\or P\or Q\or
R\or S\or T\or U\or V\or W\or X\or Y\or Z\fi}
\def\eqn#1{ \ifnum\secno>0 \eqno(\the\secno.\the\meqno)
\xdef#1{\the\secno.\the\meqno} \else\ifnum\appno>0
\eqno({\rm\romappno}.\the\meqno)\xdef#1{{\rm\romappno}.\the\meqno}
\else \eqno(\the\meqno)\xdef#1{\the\meqno} \fi \fi
\global\advance\meqno by1 }

%%% Macro to do the refs %%%%%%%%%%%%%%%%%%%%%%%%%%%%%
\global\newcount\refno \global\refno=1 \newwrite\reffile
\newwrite\refmac \newlinechar=`\^^J \def\ref#1#2%
{\the\refno\nref#1{#2}} \def\nref#1#2{\xdef#1{\the\refno}
\ifnum\refno=1\immediate\openout\reffile=refs.tmp\fi
\immediate\write\reffile{\noexpand\item{[\noexpand#1]\ }#2\noexpand%
\nobreak.} \immediate\write\refmac{\def\noexpand#1{\the\refno}}
\global\advance\refno by1} \def\semi{;\hfil\noexpand ^^J}
\def\nl{\hfil\noexpand ^^J} \def\refn#1#2{\nref#1{#2}}

\def\plA#1#2#3{{\it Phys.\ Lett.}\ {\bf A{#1}} ({#2}) #3}
\def\plB#1#2#3{{\it Phys.\ Lett.}\ {\bf B{#1}} ({#2}) #3}

\def\prA#1#2#3{{\it Phys.\ Rev.}\ {\bf A{#1}} ({#2}) #3}

\def\prD#1#2#3{{\it Phys.\ Rev.}\ {\bf D{#1}} ({#2}) #3}
\def\prl#1#2#3{{\it Phys.\ Rev.\ Lett.}\ {\bf #1} ({#2}) #3}

%%% Numbering does not start on title page %%%%%%%%%%%
\newif\iftitlepage \titlepagetrue \newtoks\titlepagefoot
\titlepagefoot={\hfil} \newtoks\otherpagesfoot \otherpagesfoot=%
{\hfil\tenrm\folio\hfil} \footline={\iftitlepage\the\titlepagefoot%
\global\titlepagefalse \else\the\otherpagesfoot\fi}

%%% Abstract %%%%%%%%%%%%%%%%%%%%%%%%%%%%%%%%%%%%%%%%%
\def\abstract#1{{\parindent=30pt\narrower\noindent\ninepoint\openup
2pt #1\par}}

%%% A nicer footnote (\note) %%%%%%%%%%%%%%%%%%%%%%%%%
\newcount\notenumber\notenumber=1 \def\note#1
{\unskip\footnote{$^{\the\notenumber}$} {\eightpoint\openup 1pt #1}
\global\advance\notenumber by 1}

%%% Date %%%%%%%%%%%%%%%%%%%%%%%%%%%%%%%%%%%%%%%%%%%%%
\def\today{\ifcase\month\or January\or February\or March\or
April\or May\or June\or July\or August\or September\or October\or
November\or December\fi \space\number\day, \number\year}

%%% More general stuff %%%%%%%%%%%%%%%%%%%%%%%%%%%%%%%
\def\pagewidth#1{\hsize= #1}  \def\pageheight#1{\vsize= #1}
\def\hcorrection#1{\advance\hoffset by #1}
\def\vcorrection#1{\advance\voffset by #1}

%%% Output layout of the text %%%%%%%%%%%%%%%%%%%%%%%%
\pageheight{23cm}
\pagewidth{15.7cm}
\hcorrection{-1mm}
\magnification= \magstep1
\parskip=5pt plus 1pt minus 1pt
\tolerance 8000
\def\bsk{\baselineskip= 15pt plus 1pt minus 1pt}
\bsk

%%% Definition of extra symbols R, C, Z, N %%%%%%%%%%%
\font\extra=cmss10 scaled \magstep0  \setbox1 = \hbox{{{\extra R}}}
\setbox2 = \hbox{{{\extra I}}}       \setbox3 = \hbox{{{\extra C}}}
\setbox4 = \hbox{{{\extra Z}}}       \setbox5 = \hbox{{{\extra N}}}

          % Actual special symbol: R

   % Actual special symbol: C

           % Actual special symbol: Z

            % Actual special symbol: N
                                                        
%%% Some useful macros %%%%%%%%%%%%%%%%%%%%%%%%%%%%%%%
\def\frac#1#2{{#1\over#2}}

\def\ket#1{|#1\rangle}

\def\pmb#1{\setbox0=\hbox{$#1$} \kern-.025em\copy0\kern-\wd0
    \kern.05em\copy0\kern-\wd0 \kern-.025em\raise.0433em\box0 }

\def\ve{\vfill\eject}

\def\({\left(}
\def\){\right)}
\def\[{\left[}
\def\]{\right]}

%Definitions in this article

\def\ket#1{|#1\rangle}

\def\({\left(}
\def\){\right)}

\def\bb{{\svec b}}
\def\ba{{\svec a}}

\def\bb{{\svec b}}

\def\bb{{\svec b}}

%Definitions of graphics%%%%%%%%%%
\def\figurea{\epsffile{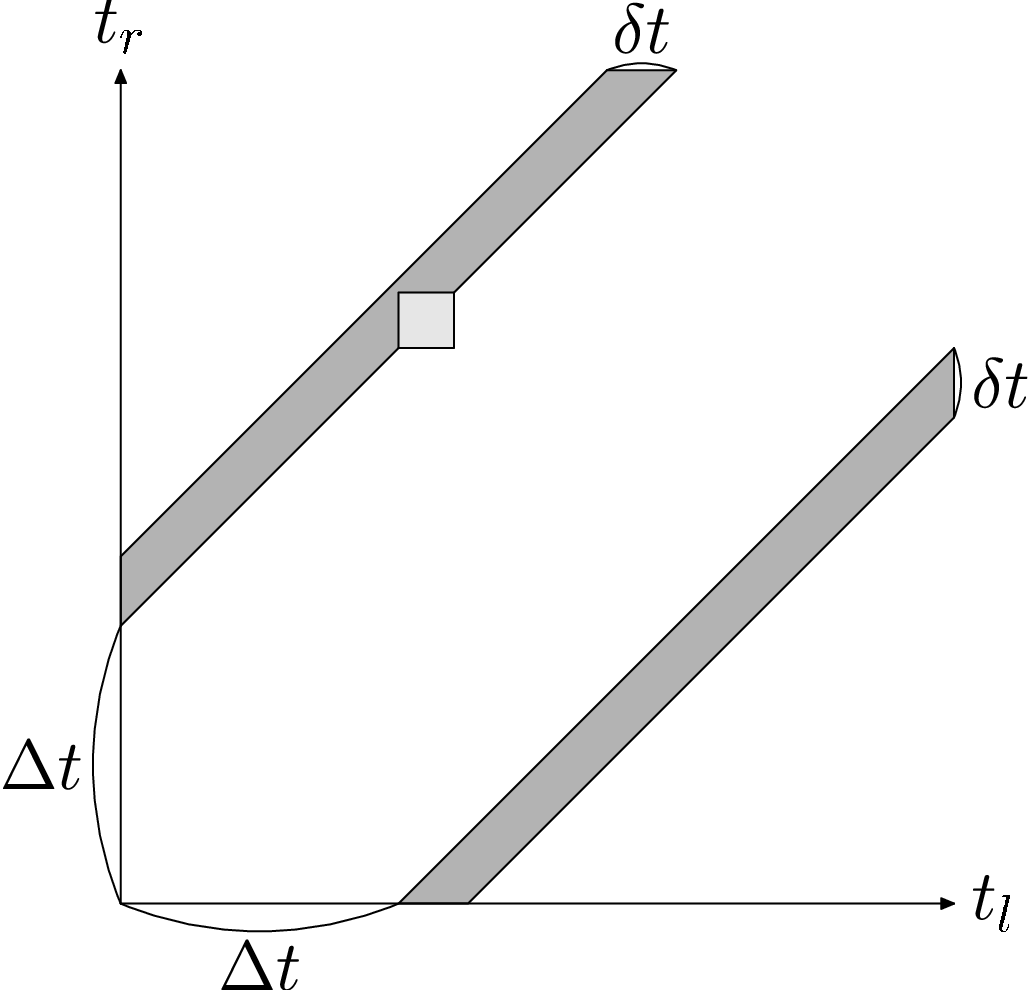}}
\def\figureb{\epsffile{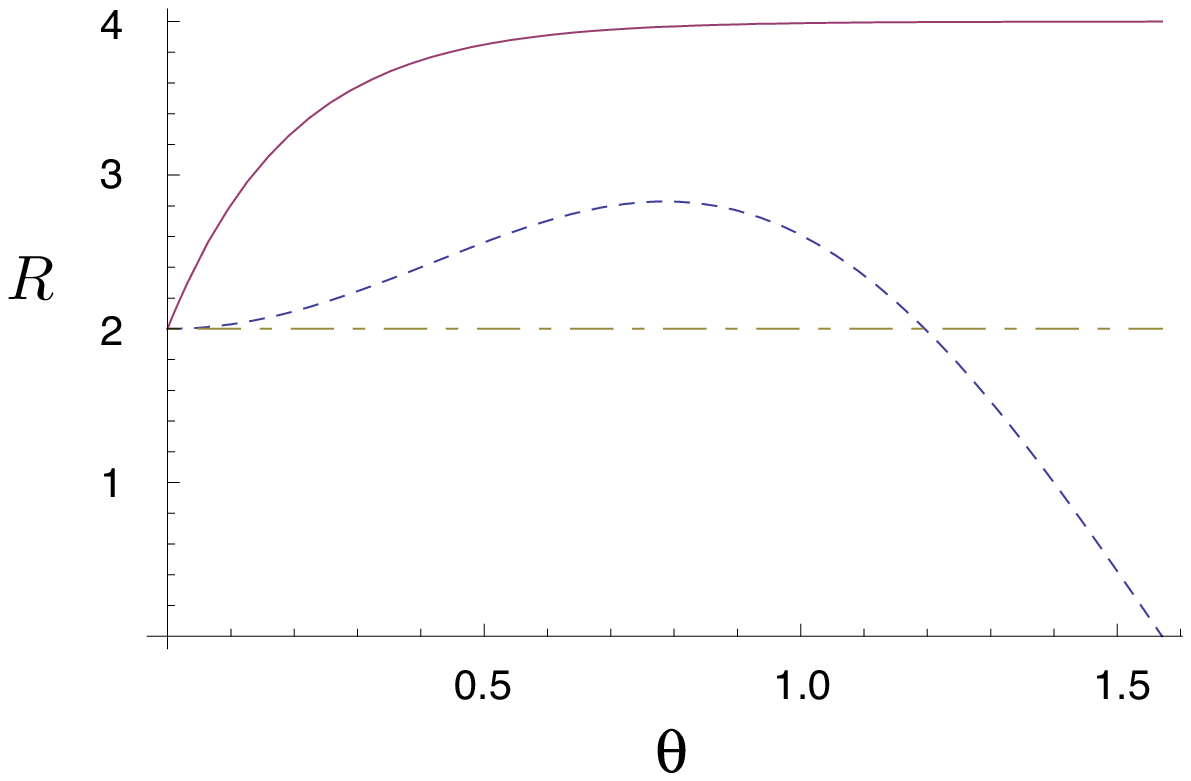}}
\def\figurec{\epsffile{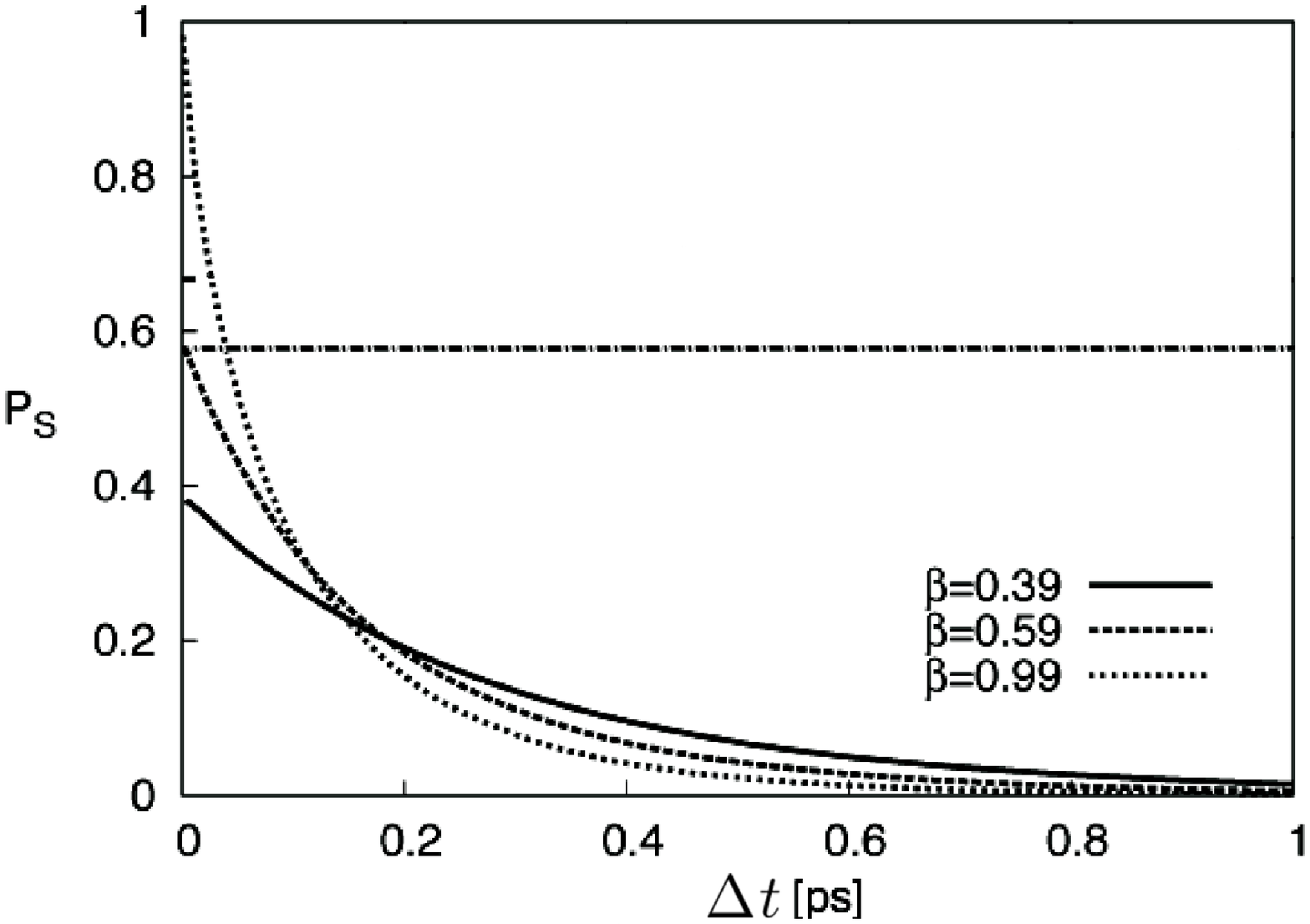}}

\input epsf.tex

%% References for the paper:

{

\refn\EPR
{A. Einstein, P. Podolsky and N. Rosen, {\it Phys. Rev.}
{\bf 47} (1935) 777}

\refn\Bellone
{J. S. Bell, {\it Physics} {\bf 1} (1964) 195}

\refn\CHSH
{J. F. Clauser, M. A. Horne, A. Shimony and R. A. Holt, 
\prl{23}{1969}{880}}

\refn\Aspect
{A. Aspect, J. Dalibard and G. Roger, \prl{49}{1982}{1804}}

\refn\Weihs
{G. Weihs, T. Jennewein, C. Simon, H. Weinfurter and A. Zeilinger, \prl{81}{1998}{5039}}

\refn\Rowe
{M. Rowe, D. Kielpinski, V. Meyer, C. Sackett, W. Itano, C. Monroe and D. Wineland,
{\it Nature}  {\bf 409} (2001) 791}

\refn\Sakai
{H. Sakai, {\it et al.}, \prl{97}{2006}{150405}}

\refn\Belltwo
{J. S. Bell, in {\it Foundations of Quantum Mechanics}, Proc. Int. Sch. of Physics, \lq Enrico Fermi\rq, ed. B. d'Espagnat, (1971) 11}

\refn\Lipkin
{H. Lipkin, {\it Phys. Rev.}
{\bf 176} (1968) 1715}

\refn\GGW
{G. C. Ghirardi, R. Grassi, and T. Weber, in {\it Proceedings of the 
Workshop on Physics and Detectors for Da$\Phi$ne}, ed. G. Pancheri, (1991) 261}

\refn\CPLEAR
{A. Apostolakis, {\it et al.} (CPLEAR Collaboration), \plB{422}{1998}{339}}

\refn\KLOE
{F. Ambrosino, {\it et al.} (KLOE Collaboration), \plB{642}{2006}{315}}

\refn\Belle
{A. Go, {\it et al.} (Belle Collaboration), \prl{99}{2007}{131802}}

\refn\Go
{A. Go, {\it J. Mod. Optics} {\bf 51} (2004) 991}

\refn\BBGH
{R. A. Bertlmann, A. Bramon, G. Garbarino and B. C. 
Hiesmayr,\plA{332}{2004}{355}}

%\refn\Leggett
%{A. J. Leggett, {\it Foundations of Physics} {\bf 33} (2003) 1469}

%\refn\Groeblacher
%{S. Gr{\"o}blacher, {\it et al.},
%{\it Nature}  {\bf 446} (2007) 871}

\refn\BEGone
{A. Bramon, R. Escribano and G. Garbarino, {\it J. Mod. Optics} {\bf 52} (2005) 1681}

\refn\Bertlmann
{R. A. Bertlmann, 
{\it Lect. Notes Phys.} {\bf 689} (2006) 1}

\refn\GG
{N. Gisin and A. Go, {\it Am. J. Phys.} {\bf 69} (2001) 264}

\refn\Santos
{Emilio Santos, quant-ph/0703206}

\refn\GM
{A. Garg and N. D. Mermin, \prD{35}{1987}{3831}}

\refn\Larsson
{J. Larsson, \prA{57}{1998}{3304}}

}

%%% Output of frontpage %%%%%%%%%%%%%%%%%%%%

\pageheight{23cm}
\pagewidth{15.7cm}
\hcorrection{0mm}
\magnification= \magstep1
\def\bsk{%
\baselineskip= 16.8pt plus 1pt minus 1pt}
\parskip=5pt plus 1pt minus 1pt
\tolerance 6000

%\vsize 19.2cm   \voffset -1.2cm   %% For temporary purposes !!!

%%%%%%%%%%%%%%%%%%%%%%%%%%%%%%%%%%

% frontpage

\hfill 
\vskip -4pt 
\hfill 
\phantom{quant-ph/0207xxx}

\vskip 42pt

%%% Setting of the baselineskip for frontpage
{\baselineskip=18pt
%%%

\centerline{\bigbold
Testing the EPR Locality using {\bigms 
B}$\,$-Mesons}

\vskip 30pt

\centerline{\smc
Tsubasa Ichikawa\footnote{${}^{*}$}
{\eightpoint email:\ tsubasa@post.kek.jp},
%\quad
%{\rm and}
\quad
Satoshi Tamura
\quad
{\rm and}
\quad
Izumi Tsutsui\footnote{${}^\dagger$}
{\eightpoint email:\ izumi.tsutsui@kek.jp}
}

\vskip 15pt

{
\baselineskip=13pt
\centerline{\it
Institute of Particle and Nuclear Studies}
\centerline{\it
High Energy Accelerator Research Organization (KEK)}
\centerline{\it Tsukuba 305-0801}
\centerline{\it Japan}
}

\vskip 125pt

\abstract{%
{\bf Abstract.}\quad
We study the possibility of testing local realistic theory (LRT), envisioned implicitly by Einstein, Podolsky and Rosen in 1935,
based on the Bell inequality for the correlations in the decay modes of entangled $K$ or $B$-mesons.   
It is shown that such a test is possible for a restricted
class of LRT, despite the passive nature of decay events and/or the non-unitary treatment of the correlations
which invalidate the test for general LRT.  Unfortunately, the present setup of the KEKB (Belle) experiment, where the coherence of entangled $B$-mesons has been confirmed recently, does not admit such a test due to the inability of determining the decay times of the entangled pairs separately.   The indeterminacy also poses a problem for ensuring the locality of the test, indicating that improvement to resolve the  indeterminacy is crucial for the test of LRT.
}
}

\vskip 10pt
\ve

%%% Output layout of the text %%%%%%%%%%%%%%

\pageheight{23cm}
\pagewidth{15.7cm}
\hcorrection{-1mm}
\magnification= \magstep1
\def\bsk{%
\baselineskip= 15.2pt plus 1pt minus 1pt}
\parskip=5pt plus 1pt minus 1pt
\tolerance 8000
\bsk

%%%%%%%%%%%%%%%%%%%%%%%%%%%%%%%%%%%%%%%%%%%%

\secno=1 \meqno=1

%%%%%%%%%%%%%%%%%%%%%%%%%%%%%%%%%%%%%%%%%%%%%%%%%%%%%%%%%%%%%%%%%%%%%
\bigskip
\noindent{\bf 1. Introduction}
\medskip
%%%%%%%%%%%%%%%%%%%%%%%%%%%%%%%%%%%%%%%%%%%%%%%%%%%%%%%%%%%%%%%%%%%%%

Entanglement lies at the heart of the recent development of quantum information theory, and yet it remains the most significant physical property in quantum mechanics (QM) that defies our intuitive understanding.  
As Einstein, Podolsky and Rosen argued in their seminal paper [\EPR], entanglement admits outcomes of observations made at two locations separated remotely to be correlated, casting an ontological question on the physical quantities observed.
Since the discovery of the Bell inequaility [\Bellone] in 1964, together with the subsequent work of Clauser,  Horne, Shimony and  Holt [\CHSH], it has been recognized that the local realistic theory (LRT) supposedly envisioned in [\EPR] can be put to test in laboratory, and  a variety of attempts for the test have been made with different entangled sources, such as photons [\Aspect, \Weihs], ions [\Rowe] and protons [\Sakai].   Almost all of the tests conducted so far indicate that non-local correlations do exist precisely as QM predicts, and we are naturally led to deny the LRT as an underlying fundamental theory, despite that these tests are still not absolutely conclusive due to their possible loopholes in the locality and detection efficiency. 

The original idea of using entangled $K$-mesons  for the test of LRT, generated as a pair $K^0$ and $\bar K^0$ via a decay in high energy experiments,  can be traced back to Bell [\Belltwo]  (and also [\Lipkin]), and it has later been elaborated by, {\it e.g.},  Ghirardi {\it et al.}[\GGW].   Part of the interest in the test with mesons derives perhaps from the curiosity as to whether the nonlocal nature of the entangled states, confirmed earlier by photons, can be extended to massive particles (for instance, the mass of a $B$-meson is more than 5 ${\rm GeV}/c^2$).
In the last ten years, experiments involving entangled pair of neutral mesons have been performed: first at CERN (CPLEAR) [\CPLEAR] and then Frascati (KLOE)  [\KLOE] with $K$-mesons, and more recently at KEK (Belle)  [\Belle] with $B$-mesons.  These experiments have confirmed that the coherent superposition of the pair of two meson states in QM is actually appropriate to describe the correlations measured for the decay modes.   However, these are not  the test of LRT, since testing LRT requires outcomes that cannot be explained by LRT, and usually this is examined by the violation of the Bell (or CHSH) inequality [\Bellone, \CHSH] obeyed by correlations of measurement outcomes based on LRT.   An argument for the possibility of such a test at 
the $B$-factories has been presented in [\Go] based on the normalized correlations of decay modes, but it has been pointed out by Bertlmann {\it et al.} [\BBGH] that this is untenable due to the passive nature of the decays and/or the non-unitary treatment of the correlations.   In fact, it has been recognized earlier in  [\GGW] (and also in [\Belltwo]) that mesons have the particular problems associated with the decays which are absent in the conventional tests using photons.

In this paper we show that, although the statements of [\BBGH]  are valid, it is still possible to carry out a test with entangled mesons, if it is designed for excluding a restricted class of LRT.   The restriction concerns with the homogeneity of decay outcomes in time and the independence of decay times from the partners in the pair decays.
%\note{%
%These requirements are formally similar to those considered for the class of nonlocal theories studied in [\Leggett, \Groeblacher].
%}
We shall see that, 
unfortunately, with the present setup of the Belle experiment such a test 
is not viable with $B$-mesons as the individual decay times are not measured; this also poses a problem for the locality
loophole.   Our result suggests that for the meaningful test of LRT at Belle, it is crucial to improve on the setup to ensure the determinacy of decay times.

\secno=2 \meqno=1

%%%%%%%%%%%%%%%%%%%%%%%%%%%%%%%%%%%%%%%%%%%%%%%%%%%%%%%%%%%%%%%%%%%%%
\bigskip
\noindent{\bf 2.  Possibility of LRT tests using meson pairs}
\medskip
%%%%%%%%%%%%%%%%%%%%%%%%%%%%%%%%%%%%%%%%%%%%%%%%%%%%%%%%%%%%%%%%%%%%%

We begin by recalling the experiment to measure the flavor of two mesons generated as an entangled
pair.  In the case of $B$-meson, for example, this can be 
realized by the process $\Upsilon(4S)\rightarrow B^0\bar{B}^0$ which is produced by the collisions of $e^-$ and $e^+$.  (The case of $K$-meson can be argued analogously, and we shall mention it when the difference becomes significant.)  
The actual measurement of flavor is carried out by looking at a particular series of 
decay modes of the mesons ($B^0 \to D^{*-} l^+ \nu$, $D^{*-} \to \bar{D^0} \pi^-_s$ and so on, with charge conjugate modes for $\bar{B^0}$), one on the left and the other on the right in the center-of-mass frame.  Due to
the flavor oscillation, the outcome of the measurement depends on the decay times of the two mesons.   Let $t_l$ and $t_r$ be the decay times of the left and right mesons, respectively.
The total ensemble of decay events of 
meson pairs for which the measurement is performed can be decomposed into subensembles 
${\cal E}(t_l, t_r)$ consisting of decay events occurring in the time cells
determined by $\{t_l, t_r\}$ and $\{t_l + \delta t, t_r+ \delta t\}$ with 
some time length $\delta t$.   To each 
subensemble ${\cal E}(t_l, t_r)$ we may consider
the correlation of the measurement outcomes of flavor for the two mesons. 

In concrete terms, one associates the value $+1$ to $B^0$ and $-1$ to $\bar B^0$ in the outcome of the measurements, and when one finds $n_{i,j}(t_l,t_r)$ decays  for the four possible outcomes $i, j$ = $\pm 1$, one evaluates the correlation in the subensemble ${\cal E}(t_l, t_r)$ by
$$
\eqalign{
C(t_l, t_r) 
&=\frac{\sum_{i,j} ij\cdot n_{i,j}(t_l,t_r)}{\sum_{i,j} n_{i,j}(t_l,t_r)} \cr
&= \frac{n_{1,1}(t_l,t_r) -n_{1,-1}(t_l,t_r) -n_{-1,1}(t_l,t_r) +n_{-1,-1}(t_l,t_r) }
{n_{1,1}(t_l,t_r) +n_{1,-1}(t_l,t_r) +n_{-1,1}(t_l,t_r) +n_{-1,-1}(t_l,t_r)}.
}
\eqn\colone
$$
Let $\Gamma$,  $\bar \Gamma$ be the decay rate of $B^0$, $\bar B^0$ to the particular modes, related by CP conjugation, which we look at in order to identify the types of mesons
in our measurement.   Ignoring the small CP-violation of the weak interaction (which is ${\cal O}(10^{-4})$ or less), we have  $\Gamma = \bar \Gamma$.
If  the total number $N$ of pairs produced in the 
experiment is sufficiently large, the number of decays belonging to the subensemble 
${\cal E}(t_l,t_r)$ is given by
$$
n_{i,j}(t_l,t_r) = N \, (\delta t)^2\, 
\Gamma_{i,j}(t_l, t_r),
\eqn\nN
$$
where $\Gamma_{i,j}(t_l, t_r)$ is the joint decay rate 
$$
\Gamma_{i,j}(t_l, t_r) =P_{i,j}(t_l,t_r)\, \Gamma^2
\eqn\jdr
$$
expressed in terms of the joint probability $P_{i,j}(t_l,t_r)$ of mesons possessing the 
flavor $i$ and $j$ at $t_l$ and $t_r$, respectively [\BEGone]. 
Combining (\nN) and  (\jdr) with (\colone), we 
find
$$
C(t_l, t_r) =\frac{\sum_{i,j} ij\cdot P_{i,j}(t_l,t_r)}{\sum_{i,j} P_{i,j}(t_l,t_r)}.
\eqn\coltwo
$$

Now that the 
the correlation is given by the joint probability $P_{i,j}(t_l,t_r)$, one may evaluate it in QM as
$$
C_{\rm Q}(t_l,t_r)=-\cos(\Delta m(t_l-t_r)),
\eqn\asymQM
$$
where $\Delta m$ is the mass difference between the weak eigenstates 
$\ket{B_L} = (\ket{B^0} + \ket{\bar B^0})/\sqrt{2} $ and $\ket{B_H} = (\ket{B^0} 
- \ket{\bar B^0})/\sqrt{2}$ (see, {\it e.g.}, [\Bertlmann]). 
The correlation (\asymQM) in QM violates the Bell inequality, provided that there exists an
LRT describing the meson system in which the decay times $(t_l,t_r)$ play a similar role as the angle parameters in the usual photon polarization experiment.  Since the angle parameters can be adjusted freely by the observer while decay times cannot, we need to examine the feasibility of such a test closely.

An LRT for the system of  meson pairs may generally be formulated based on the following premises.  First, the 
theory has a set of hidden parameters, collectively denoted by $\Lambda$, which determines the physical 
states of the system completely (here we consider only deterministic 
LRT, but stochastic LRT can be dealt with analogously).   Second, it is 
equipped with a probability distribution $\xi(\Lambda)$ which is a 
non-negative function of $\Lambda$ normalized as
$\int d\Lambda\, \xi(\Lambda)=1$.  

Let $A({\ba})$ (and $B({\bb})$) be an observable of the left (and right) particle in the decayed 
pair, with ${\ba}$ (and ${\bb}$) being some external parameter specifying the measurement 
setup of the observable.   
In LRT,  the outcome of a measurement of $A({\ba})$ is determined 
by $\Lambda$,  and by locality requirement it is independent of the external parameter ${\bb}$ or the outcome of measurement made on the right particle.   This allows us to write the outcome of the measurement of $A({\ba})$ as $A(\Lambda, {\ba})$.  Similarly, the outcomes of the measurement for the right particle is denoted by $B(\Lambda, {\bb})$.  If these observables are dichotomic and yield only 
$\pm 1$ values, say, $+1$ for  $B^0$ and $-1$ for  $\bar B^0$ to be consistent with the previous assignment of the $B$-meson experiment, then
the correlation of the outcomes of the joint measurement reads
$$
{\cal C}_{\rm L}(\ba, \bb) =\int d\Lambda\, \xi(\Lambda)\, 
A(\Lambda, {\ba})B(\Lambda, {\bb}).
\eqn\clrt
$$

In the experiments attempting to test the Bell inequality using photons,
we measure the spin (polarization) in arbitrary directions we choose, which act as external parameters represented by 
vectors ${\ba}$ and ${\bb}$.  In the case of  meson pairs,
we measure the flavor 
of the particles in a specific direction determined by $B^0$ and $\bar B^0$ so that they can be 
clearly distinguished from the decay modes.   Here lies a
salient feature of the LRT test based on the Bell inequality using meson pairs:
the direction of the flavor measurement is 
fixed in the flavor space while the 
actual flavor of the particles changes in time due to the $B^0$-$\bar B^0$ 
oscillation.   
This is to be contrasted to the case of photons where the spin is measured 
in any directions while the spin of the photon is fixed in time. \note{%
It is possible, however, to put the two cases in a unified framework in which, {\it e.g.}, birefringence of photons corresponds to the oscillation in mesons [\GG].}
The analogy of the two cases is nevertheless recognized 
if we notice that the decay times
$t_l$ and $t_r$ of the left and right particle can be regarded, at least formally, as parameters corresponding to ${\ba}$ and ${\bb}$ on account of the oscillation in the flavor space.
This suggests that the Bell inequality may also hold for the correlation (\clrt) of the meson decays 
with ${\ba}$ and ${\bb}$ replaced by $t_l$ and $t_r$, for which the test could be performed as in the photon case. 

There is, however, a crucial difference between the two cases.  That is, since the 
decay times  $\{t_l, t_r\}$ cannot be
adjusted freely in the actual measurement
(carried out at KLOE for $K$-mesons and at KEK for $B$-mesons),
they are of passive character and, as such, should be more properly
regarded as part of the hidden parameters $\Lambda$ rather than external parameters.   
It follows that the correlation (\clrt) has a parameter-dependent probability 
distribution $\rho(\Lambda)$, and the Bell inequality cannot be derived there.

To analyze this \lq passiveness problem\rq\ further, let us put 
$\Lambda = \{\lambda, t_l, t_r\}$ with $\lambda$ representing the rest of the hidden parameters\note{% 
The space of the hidden parameters $\Lambda$, or the subspace of $\lambda$ in it, could be topologically nontrivial, but the following arguments are not affected by this as long as the independence of $\lambda$ from 
$t_l, t_r$ is ensured. 
} 
in the LRT.    
To each 
subensemble  ${\cal E}(t_l, t_r)$ with sufficiently small span $\delta t$ of the time cell, the 
correlation deduced from (\clrt) is given by
$$
C_{\rm L}(t_l, t_r)
=\int d\lambda\, {\rho}(\lambda\,\vert\,  t_l, t_r)\,A(\lambda, t_l, 
t_r)B(\lambda, t_l, t_r),
\eqn\clrttwo
$$
where we have used the density for the subensemble,
$$
{\rho}(\lambda \,\vert\,  t_l, t_r) := {{{\xi}(\Lambda)}\over{\int 
d\lambda\, {\xi}(\Lambda)}},
\eqn\subden
$$
which is normalized as
$\int d\lambda\, \rho(\lambda \vert  t_l, t_r) = 1$.  
In (\clrttwo) we have omitted the external parameters ${\ba}$ and ${\bb}$ which 
are absent in the experiment under consideration, and instead 
made explicit the dependence on $t_l$ and $t_r$.   As mentioned 
above, the expression (\clrttwo) shows that, when $\{t_l, t_r\}$ are 
interpreted as external parameters, the deduced LRT describing the subensemble ${\cal E}(t_l, t_r)$ is 
highly nonlocal and hence cannot lead to the standard Bell inequality.  

Although we seem to be forced to give up testing the general LRT for the meson system, we may still
seek conditions under which the correlation (\clrttwo)
takes the standard form  (\clrt).  For this to be the case, an obvious condition is that the density (\subden) of the subensemble be independent of $t_l$ and $t_r$.  This is ensured if the total density 
${\xi}(\Lambda)$ decouples as
$$
{\xi}(\Lambda) = {\rho}(\lambda)\,{\eta}(t_l, t_r),
\eqn\deccon
$$
where both of the factors ${\rho}(\lambda)$, ${\eta}(t_l, t_r)$ are normalized with respect to the arguments (which can always be done without loss of generality).
Indeed, this ensures the density (\subden) to be independent of the subensemble, ${\rho}(\lambda \,\vert\,  t_l, t_r) = {\rho}(\lambda)$.  Thus, 
the condition (\deccon) states that, in effect, all the subensembles ${\cal E}(t_l, 
t_r)$ are identical as a probability set in that they share the same normalized probability density ${\rho}(\lambda)$.   The only difference is found in the scaling factor ${\eta}(t_l, t_r)$ which describes the variation in the number of decay events in each of the subensembles ${\cal E}(t_l, t_r)$, which decreases for larger $t_l$ and $t_r$.
One may argue that the condition (\deccon) is physically plausible from the viewpoint of homogeneity in time of the decaying phenomena, and it can surely be
examined by observing the time dependence of the decay modes.   
%Note that the outcomes of the measurements can nevertheless depend on $\{t_l, t_r\}$ if the observables are themselves dependent on the decay times.

Another condition required is that the outcomes of the measurements performed on 
one particle be independent of the decay time of the other particle which 
are spatially separated from  each other:
$$
A(\lambda, t_l, t_r) = A(\lambda, t_l),  \qquad 
B(\lambda, t_l, t_r)=B(\lambda, t_r).
\eqn\indcon
$$
Note that this is not the conventional locality condition, because our LRT possesses
$\{t_l, t_r\}$ as part of the hidden parameters $\Lambda$.  
%It becomes the conventional one only when $\{t_l, t_r\}$ are treated as external parameters.
In words, the second condition (\indcon) states that the outcome of measurement 
may depend on the decay time of the measured particle  but it is independent of the decay time of the other particle in a remote distance.  
Combining the  independence condition (\indcon) for decay time with the 
homogeneity condition (\deccon), one finds that the correlation (\clrttwo) 
reduces to
$$
C_{\rm L}(t_l, t_r)
=\int d\lambda\, {\rho}(\lambda)\,A(\lambda, t_l)\, B(\lambda, t_r).
\eqn\corsub
$$

To see that the correlation  (\corsub) corresponds to the quantity evaluated experimentally, 
we introduce the \lq step\rq\ functions
$\Theta_i^A(\lambda, t_l)$, $i = \pm 1$,  which determine the flavor type of the left meson such that $\Theta_1^A(\lambda, t_l) = 1$ if the type is $B^0$ and $\Theta_1^A(\lambda, t_l) = 0$ if it is $\bar B^0$, and conversely 
$\Theta_{-1}^A(\lambda, t_l) = 0$ if it is $B^0$ and $\Theta_{-1}(\lambda, t_l) = 1$ if it is $\bar B^0$. 
With analogously defined functions $\Theta_i^B(\lambda, t_r)$ for the right meson, we can write the values of the observables  as
$$
A(\lambda, t_l) = \Theta_1^A(\lambda, t_l) - \Theta_{-1}^A(\lambda, t_l),  \qquad 
B(\lambda, t_r) = \Theta_1^B(\lambda, t_r) - \Theta_{-1}^B(\lambda, t_r).
\eqn\abex
$$
Plugging (\abex) into (\corsub), and using the identities 
$\sum_i \Theta_i^A(\lambda, t_l) = \sum_i \Theta_i^B(\lambda, t_r) = 1$, we find 
$$
C_{\rm L}(t_l, t_r) =\frac{\sum_{i,j} ij\cdot P^{\rm L}_{i,j}(t_l,t_r)}{\sum_{i,j} P^{\rm L}_{i,j}(t_l,t_r)},
\eqn\collrtsp
$$
where
$$
P^{\rm L}_{i, j}(t_l, t_r)
=\int d\lambda\, {\rho}(\lambda)\, {\eta}(t_l, t_r) \,\Theta_i^A (\lambda, t_l) \,  \Theta_j^B (\lambda, t_r)
\eqn\jointp
$$
is the joint probability of finding the mesons in flavors $i$ and $j$ at times $t_l$ and $t_r$, respectively, in the LRT.
This establishes the link between our LRT and the actual experiments in the correlations (\collrtsp) and (\coltwo), or  (\corsub) and  (\colone).  

Now that the formal structure of the correlation  (\corsub) is exactly the same as the standard one, 
we have
$$
\vert C_{\rm L}(t_l, t_r) + C_{\rm L}(t_l', t_r) + C_{\rm L}(t_l', t_r') - C_{\rm L}(t_l, t_r') \vert \le 2,
\eqn\bellineq
$$
which is equivalent to the Bell inequality.  It follows that, for the restricted class of  LRT which fulfill the above two conditions, the correlations defined for the subensumbles ${\cal E}(t_l, t_r)$ satisfy the Bell inequality (\bellineq), despite that  $t_l$ and $t_r$ are not external parameters.  
Consequently, we see that as assumed in [\Go] the correlation (\colone) evaluated from the 
experiment can indeed be used to test the restricted class of LRT based on the Bell inequality  (\bellineq) which is violated by the QM correlations (\asymQM) for $B$-mesons.   This is also the case with the $K$-mesons, although we have an extra damping factor for the quantum correlations (\asymQM) due to the large difference in the lifetimes of $K_L$ and $K_S$ rendering the violation less evident [\GG].   For these cases, seeing violation of the inequality in meson pair decays allows us to reject any LRT which fulfills the conditions  (\deccon) and  (\indcon).

Before examining the feasibility of LRT test in experiments,  it is worth considering a special case of the class of LRT in which a rather simple picture of local realism can be realized.  The case arises when the  density ${\eta}(t_l, t_r)$ for the decay times decouples into two densities, 
$$
{\eta}(t_l, t_r) = {\eta}_l(t_l)\, \eta_r( t_r),
\eqn\no
$$
where ${\eta}_l(t_l)$ and $\eta_r( t_r)$ depend only on the decay times of the respective 
mesons. The normalization $\int_0^\infty dt_l {\eta}_l(t_l) = \int_0^\infty dt_r {\eta}_r(t_r) = 1$ assured in (\deccon)  suggests that ${\eta}_l(t_l)$ and $\eta_r( t_r)$ may be regarded as the decay probability densities of the individual mesons.   From these,  
one can form the products of the decay probability densities and the functions specifying the flavor types to obtain the decay probability densities separately for the two mesons,
$$
P_i^A (\lambda, t_l) = {\eta}_l(t_l) \, \Theta_i^A(\lambda, t_l), \qquad
P_i^B (\lambda, t_r) = {\eta}_r(t_r) \, \Theta_i^B(\lambda, t_r).
\eqn\prbden
$$
The joint probability (\jointp) can then be expressed in terms of the individual  decay probability densities,
$$
P^{\rm L}_{i, j}(t_l, t_r)
=\int d\lambda\, {\rho}(\lambda)\, P_i^A (\lambda, t_l) \,  P_j^B (\lambda, t_r),
\eqn\no
$$
in which the independence in the decays, which is perhaps natural from the locality point of view, is ensured.
Note that in the present case the split form of the local probability densities (\prbden) implies
$$
\sum_{i = \pm 1} P_i^A (\lambda, t_l) = {\eta}_l(t_l), \qquad \sum_{i = \pm 1} P_i^B (\lambda, t_r) = {\eta}_r(t_r).
\eqn\lrtcond
$$
We mention that the LRT model presented in [\Santos] which yields the quantum correlation (\asymQM) does not fulfill (\lrtcond) nor even the the homogeneity condition (\deccon),  which is also the case with the example of LRT mentioned in [\BEGone].

\secno=3 \meqno=1

%%%%%%%%%%%%%%%%%%%%%%%%%%%%%%%%%%%%%%%%%%%%%%%%%%%%%%%%%%%%%%%%%%%%%
\bigskip
\noindent{\bf 3. Possibility of LRT test at KEKB (Belle)}
\medskip
%%%%%%%%%%%%%%%%%%%%%%%%%%%%%%%%%%%%%%%%%%%%%%%%%%%%%%%%%%%%%%%%%%%%%

Now we turn to the question of the possibility of LRT test at KEKB (Belle experiment), where entangled $B$-mesons are being produced and  a large number of data (8565 events for the measured series of modes out of $152 \times 10^6$ events) have been used for examining the entanglement in the analysis of Ref.[\Belle].  
We first point out that the subensembles introduced above are not appropriate for the present setup of the Belle experiment 
and should be replaced by a modified set of subensembles for constructing LRT for which a test may be conducted. 

In the Belle experiment, one measures the distance $\Delta z$ of decay points for each pair along the direction of the beam defined as anti-parellel to the positron beam line.   Since the velocities of the $B$-mesons are negligible compared to the velocity of the $\Upsilon(4S)$ which has $\beta \gamma = 0.425$, 
the proper time difference $\Delta t =  \vert  t_l-t_r \vert$ can be estimated from $\Delta z$ by 
$\Delta t \approx \Delta z/\beta\gamma c$.   An important point here is that, due to the large uncertainty on the $\Upsilon(4S)$ decay points, we only have $\Delta t$, not the separate values $t_l$ and $t_r$,  and hence we cannot reconstruct  the subensemble ${\cal E}(t_l, t_r)$ from the data.   This forces us to consider a different set of subensembles ${\cal E}(\Delta t)$ characterized by the time difference $\Delta t$ only.  Namely, we introduce subensembles 
${\cal E}(\Delta t)$ consisting of decay events occurring in the two time strips defined by the areas between the lines $\{t_l, t_r\} = \{t, t\pm\Delta t\}$ and $\{t +\delta t, t\pm\Delta t+ \delta t\}$ with 
some time length $\delta t$ for all $t \ge 0$ (see Fig.~1).    The outcomes of observation for the mesons in the pair then define the corresponding correlation $C(\Delta t)$ specified by the decay time difference $\Delta t$.

%%%%%%%%%%%%%%%%%
\topinsert   
\centerline{
\epsfxsize=2in
\epsfysize=1.8in
\figurea
}
\bigskip
\abstract{{\bf Figure 1.} 
The subensemble ${\cal E}(\Delta t)$ consists of two strips in the plane of decay times $(t_l, t_r)$ specified by the time difference $\Delta t =  \vert  t_l-t_r \vert$.  The subensemble ${\cal E}(t_l, t_r)$ is shown by the box on one of the strips.}
\bigskip
\endinsert
%%%%%%%%%%%%%%%%%

To find a proper expression for the correlation in a LRT, we may again assume the conditions  (\deccon) and (\indcon) restricting the LRT.  Then the correlation in the outcomes associated with the subensemble ${\cal E}(\Delta t)$ is given by
$$
\eqalign{
C_{\rm L}(\Delta t)
&= {1\over{N(\Delta t)}}
\int d\lambda\, {\rho}(\lambda)\,\int_0^\infty dt\, \bigl\{\eta(t +\Delta t, t)  A(\lambda, t +\Delta t)B(\lambda, t) \cr
&\qquad \qquad \qquad\qquad \qquad\qquad   + \eta(t, t +\Delta t)  A(\lambda, t)B(\lambda,  t +\Delta t)\bigr\},
}
\eqn\corbellesub
$$
where $N(\Delta t)$ is a normalization factor,
$$
N(\Delta t) = \int_0^\infty dt\, \bigl\{\eta(t +\Delta t, t) + \eta(t, t +\Delta t) \bigr\},
\eqn\no
$$
which ensures that $C_{\rm L}(\Delta t) = \pm 1$ for perfect (anti-)correlations.   

With this correlation, one may seek an inequality similar to the Bell inequality by considering the combination,
$$
R := \big\vert C_{\rm L}(\Delta t_{1{1}'}) + C_{\rm L}(\Delta t_{1{2}'}) + C_{\rm L}(\Delta t_{2{1}'}) - C_{\rm L}(\Delta t_{2{2}'})\big\vert,
\eqn\rfactor
$$
for four different time differences $\Delta t_{i{j}'}$, $i, j = 1, 2$.   The analogy with the Bell inequality is realized by associating a set of times $\{ t_1, t_{1'}, t_2, t_{2'}\}$ such that $\Delta t_{i{j}'} = \vert t_i - t_{j'}\vert$.
To proceed, we assume for simplicity that the temporal density is symmetric $\eta(t +\Delta t, t) = \eta(t, t +\Delta t)$.  We also introduce 
$$
\tilde \eta(t; \Delta t) := {1\over{N(\Delta t)}}\Theta(t)\, \eta(t, t +\Delta t),
\eqn\etfunct
$$
with the step function $\Theta(t)$, 
which is defined for all $-\infty < t < \infty $ and normalized as $\int_{-\infty}^\infty dt\,   \tilde \eta(t; \Delta t ) = 1/2$.
The factor $R$ in (\rfactor) then becomes
$$
R = \bigg\vert
\int d\lambda\, {\rho}(\lambda)\,\sum_{i,j} s_{i{j'}} \int_{-\infty}^\infty dt\, \tilde \eta(t+\tau_{i{j}'} ; \Delta t_{i{j}'} )
\bigl\{A(\lambda, t+t_i)B(\lambda, t+t_{j'}) + (i \leftrightarrow {j'})\bigr\} \bigg\vert,
\eqn\no
$$
where we have used $s_{i{j'}}$ defined by  $s_{11'} = -1$ and $s_{i{j'}} = +1$ otherwise, and $\tau_{i{j}'} = \min\{t_i, t_{j'}\}$.

To evaluate an upper bound of $R$, we choose an arbitrary function $f(t) \ge 0$ and employ the shorthand
$X_{i{j'}} := A(\lambda, t+t_i)B(\lambda, t+t_{j'})$ to obtain
$$
\eqalign{
R 
&\le  \bigg\vert
\int d\lambda\, {\rho}(\lambda)\,\sum_{i,j} s_{i{j'}} \int_{-\infty}^\infty dt\, 
\bigl\{\tilde \eta(t+\tau_{i{j}'} ; \Delta t_{i{j}'} )X_{i{j'}}  - f(t)X_{i{j'}}\bigr\}\bigg\vert \cr
&\qquad + \bigg\vert\int d\lambda\, {\rho}(\lambda)\,\sum_{i,j} s_{i{j'}} \int_{-\infty}^\infty dt\,   f(t)X_{i{j'}}\bigg\vert 
+  (i \leftrightarrow {j'}).
}
\eqn\bineqone
$$
Now, if we recall the property
$\vert \sum_{i,j} s_{i{j'}} X_{i{j'}} \vert \le 2$ which is the key element of the Bell inequality valid for dichotomic variables $X_{i{j'}}= \pm 1$, we find that the second term in the r.h.s.~of (\bineqone) is bounded by $2 \int_{-\infty}^\infty dt\, f(t)$. In order to find a better (stringent) upper bound, we may choose $f(t)$ such that $f(t) \le \tilde \eta(t+\tau_{i{j}'} ; \Delta t_{i{j'}} )$ for all $i, j$.  Using $\vert s_{i{j'}} X_{i{j'}} \vert = 1$ in the first term in the r.h.s.~of (\bineqone), 
we obtain
$$
R \le  4 - 4\int_{-\infty}^\infty dt\, f(t).
\eqn\bineqtwo
$$
If the density $\tilde \eta(t; \Delta t)$ is a monotonically decreasing function of both $t$ and $\Delta t$ (which is the 
case in the natural decaying phenomena), then the best choice for 
$f(t)$ is obviously $f(t) = \Theta(t+\tau_{\rm min})\, \tilde \eta(t+\tau_{\rm max}; \Delta \bar t)$,  where 
$\tau_{\rm min}$ and $\tau_{\rm max}$ are, respectively, the minimal value and the maximal value among $\{\tau_{1{1}'}, \tau_{1{2}'},\tau_{2{1}'},\tau_{2{2}'} \}$, and $\Delta \bar t$ is the larger one in the two time differences $\Delta t_{i{j}'}$ for which $\tau_{\rm max} = \tau_{i{j}'}$ holds. Plugging this into (\bineqtwo), we arrive at
$$
R \le  2 + 4\int_{0}^{\tau_{\rm max}-\tau_{\rm min}} dt\,  \tilde\eta(t ; \Delta \bar t).
\eqn\bineq
$$
The second term in the r.h.s.~of (\bineq) represents an increase in the upper bound for the combination of the correlations (\corbellesub) which is  larger than the value 2 of the standard Bell inequality.

%%%%%%%%%%%%%%%%%
\topinsert   
\centerline{
\epsfxsize=2.6in
\epsfysize=1.8in
\figureb
}
\bigskip
\abstract{{\bf Figure 2.} 
The factor $R$ as a function of $\theta$ under the combination $( t_1, t_{1'}, t_2, t_{2'}) = (\theta, 2\theta, 3\theta, 0)/\Delta m$.  The solid curve shows the upper bound (\bineq) for $R$ for the LRT with 
$\eta(t_l, t_r) \propto e^{-\Gamma(t_l + t_r)}$ in which $\tau_{\rm max}-\tau_{\rm min} = 2\theta/\Delta m$.  The upper bound is lifted from the standard value $2$ and is well above  the value of QM shown by the dotted curve.   The wide gap between the two curves indicates that the LRT test is not feasible with the data obtained under the present setup of the experiment.    
}
\bigskip
\endinsert
%%%%%%%%%%%%%%%%%

Unfortunately, the increased upper bound in (\bineq) is likely to invalidate our test of LRT.   To see this, recall that the QM correlation (\asymQM) depends only on the time difference and hence remains valid for the correlation we are considering in the subensemble ${\cal E}(\Delta t)$, {\it i.e.}, 
 $C_{\rm Q}(\Delta t) = C_{\rm Q}(t_l,t_r)$  for $\Delta t = \vert t_l-t_r \vert$.  Thus the largest  value attained by QM for the factor $R$ is $2\sqrt{2} \simeq 2.83$, which is realized, {\it e.g.}, by the choice $( t_1, t_{1'}, t_2, t_{2'}) = (\pi/4, \pi/2, 3\pi/4, 0)/\Delta m$.    For numerical comparison, we may adopt the density
 $\eta(t_l, t_r) \propto e^{-\Gamma(t_l + t_r)}$ (which may be confirmed from the observed decay law) and the values
 $\Delta m \simeq 5.02 \times 10^{11} s^{-1}$ and $\Gamma \simeq 6.49 \times 10^{11} s^{-1}$ for the $B$-mesons, under which 
the upper bound for in the LRT in (\bineq) is  $2 + 2(1-e^{-\pi\Gamma/\Delta m}) \simeq 3.97$.  This suggests that the LRT can account for the correlations for all values of $R$ obtained in quantum mechanics and, therefore, testing the LRT with the inequality (\bineq) is not possible.  This is illustrated in Fig.~2, where the factor $R$ is evaluated as a function of $\theta$ when we choose the combination 
$( t_1, t_{1'}, t_2, t_{2'}) = (\theta, 2\theta, 3\theta, 0)/\Delta m$.   From Fig.~2 we find that for the upper bound for $R$
for LRT to be smaller than the maximal value of QM at $\theta = \pi/4$, we need\note{%
Other values for the ratio $x$ have been mentioned in [\BBGH] as a feasibility measure of general LRT test when the passive nature of decay is neglected.  Note that the meaning of the measure is different in our discussion, and the required value of $x$ is found to be somewhat larger than the values mentioned there.
}   
$x := \Delta m/\Gamma \ge 5.9$,
which can be fulfilled neither by the
$B$-meson ($x \simeq 0.77$) nor the $K$-meson ($x \simeq 0.95$).
More generally, one can also seek other combinations for the time differences $\Delta t_{i{j}'}$ in order to examine if there are cases where the test becomes meaningful.  Our results by Monte Carlo simulation shows that there are no such cases, either.

Next we turn to the question of the locality loophole, which should be addressed for 
all experiments designed to test the LRT.
The locality loophole refers to the possibility of  
communication between the local measurements such that the external cause (tuned by the observer) or the outcome of one of the measurements -- in the present case one of the observed decay mode -- may  
influence the outcome of the other measurement.  In the restricted  class of LRT we are considering where the decay times are not regarded as external parameters, the locality is ensured if the two measurement outcomes do not depend on each other (\lq outcome independence\rq).   In the actual observation, the locality requires basically that pairs of the 
decay events be space-like separated.    Here the problem with the Belle experiment is that 
the spatial distance $\Delta z$ measured for the decay events is insufficient to tell whether the events are separated space-like or not. 
Thus, all we can do is to estimate how much the events counted in the experiment are 
space-like and pass the locality requirement.   We do this  by adopting again the assumption that the temporal density 
be of the form  $\eta(t_l, t_r) \propto e^{-\Gamma(t_l + t_r)}$ and further that the decays occur isotropically in the center-of-mass frame of the mesons.   Then the result of our computer simulation shown in Fig.~3 indicates that only a small fraction of events fulfill the locality requirement, and that the situation will not improve even if we alter the energy 
assymetry of the $e^{+}e^{-}$-collider, {\it e.g.}, from the present value 
0.375 GeV + 74.6 GeV.

%%%%%%%%%%%%%%%%%
\topinsert   
\centerline{
\epsfxsize=3.2in
\epsfysize=2.4in
\figurec
}
\bigskip
\abstract{{\bf Figure 3.} 
Probability $P_{\rm s}$ of space-like decay events as a function of 
$\Delta t$.   The solid line is for the case $\beta = 0.39$ which is the present value of the Belle experiment.  
The lower limit $P_{\rm s} = 0.59$  is barely cleared by $\beta = 0.59$ at $\Delta t = 0$, but even the choice $\beta = 0.99$ leaves only a tiny range of $\Delta t$ which is insufficient for the LRT test.
%Unfortunately, none of them has a sufficient range under which the LRT test can yield a conclusive result, as they are far away from the significant range where the upper limit  2, represeted by the horizontal line,  representing the  represents the lower limit $P_{\rm s} = 0.59$ for the test being free from the locality loophole.
}
\bigskip
\endinsert
%%%%%%%%%%%%%%%%%

To furnish a simple bound for the ratio of space-like events, 
suppose that our decays  consist of a mixture of space-like and non space-like events with the ratio $p$ to $1-p$. 
For space-like events, the correlation in our restricted LRT is given by (\corsub) for which the upper bound 
of the combination (\bellineq) is 2.  On the other hand, for non space-like events the outcome of the measurement 
$A$ can take the form $A(\lambda, t_l, B)$ where $B$ represents the outcome of the measurement of $B$, and similarly we have $B = B(\lambda, t_r, A)$.  The correlation of the outcomes then admits values up to 4 for the combination (\bellineq).  
Denoting the former (normalized) correlation by $C_{\rm L}^{\rm s}$ and the latter by $C_{\rm L}^{\rm ns}$, we see that 
the actual correlation that can be evaluated from experimental data is the mixture 
$C_{\rm L} = p\, C_{\rm L}^{\rm s} + (1-p)\, C_{\rm L}^{\rm ns}$.   With the mixed correlation, the upper bound for the factor $R$ in (\rfactor) is then 
$$
R \le2p+4(1-p).
\eqn\genineq
$$
The upper bound is less than the quantum upper bound $2\sqrt{2}$ if
$$
p>\sqrt{2}(\sqrt{2}-1)\simeq 0.59,
\eqn\lb
$$
which gives a lower bound for the ratio of space-like events in the decay events.   As we can see in Fig.~3, the ratio cannot be attained for the range of time differences required to test the LRT.

\secno=4 \meqno=1

\bigskip
\noindent{\bf 4. Conclusion and discussions}
\medskip

In this paper we argued that entangled mesons can be used for the test of LRT, if it is restricted to a class in which the decay times $\{t_l, t_r\}$ of the meson pairs can be treated formally as if they are external parameters.  
The possibility of the test has been discussed previously for $K$-mesons  [\GG] and also for $B$-mesons [\Go], both based on the normalized correlations adopting the fair sampling assumption.    The drawbacks associated with the meson experiments are known to be two-fold [\BBGH]: one is the passiveness of the decays which invalidates the treatment of the times  $\{t_l, t_r\}$ as external parameters, and the other is the dubious use of the normalized correlations for the test of the Bell inequality.   
These drawbacks render the test of LRT untenable for the most general class, but they can be avoided for the restricted class of LRT,  in which the standard Bell inequality holds formally with the normalized correlations which are well-defined in the subensembles ${\cal E}(t_l, t_r)$, even though the parameters $\{t_l, t_r\}$ are not external intrinsically. 

Unfortunately, the present setup of the Belle experiment at KEK concentrates only 
on the specification of the time difference $\Delta t = \vert t_l -  t_r\vert$ and not on the decay times $t_l$ and $t_r$ separately, and accordingly we are led to considering a larger set of subensembles ${\cal E}(\Delta t)$ consisting of ${\cal E}(t_l, t_r)$ with the same $\Delta t$.   The correlations in the new subensembles still obey an inequality analogous to the Bell inequality but with a loose upper bound compatible with the quantum value, implying that the test of LRT, even in the restricted class, cannot be done conclusively.   
The separate specification of decay times, to a certain degree of resolution, is also required to close the locality loophole, since knowing the decay times enables us to choose space-like decay events only.   Unless the specification is made possible, we find  statistically that the inevitable inclusion of time-like events lifts the upper bound of the inequality and, consequently, the test of the LRT becomes unviable.

Finally, we briefly mention the question of the efficiency loophole for the Belle experiment.  It is known  [\GM, \Larsson] that in general a test of LRT becomes inconclusive unless the detector efficiency exceeds 82.8\% for dichotomic variables.   In the case of the Belle experiment,  even though the detection rate is reasonably high for individual decays in the particular series of modes measured, the overall efficiency is possibly reduced to 10\% after completing the multiple decay processes involved.   Since the inefficiency in the detection is largely due to the angle deficit of detectors, we need to invoke a fair sampling assumption to enhance the overall detection efficiency, or otherwise we seek some other type of decay modes (which might actually require mesons other than $B$ or $K$) in which a fewer number of processes are involved.
This is certainly an issue to be studied further, but once it is cleared and the specification of decay times is made possible
along the line discussed above, the test of LRT with mesons based on the Bell inequality will become an interesting possibility for probing the basic nature of quantum mechanics with massive and presumably more localized objects than other particles used so far.

\bigskip
\noindent
{\bf Acknowledgement:}
We thank Dr.~Y.~Sakai, Dr.~B.~Yabsley and
Dr.~K.~Komatsubara
for useful comments and also for information concerning the experiments at Belle and KLOE.
This work has been supported in part by
the Grant-in-Aid for Scientific Research (C), No.~16540354 and 20540391, of
the Japanese Ministry of Education, Science, Sports and Culture.

\baselineskip= 15.5pt plus 1pt minus 1pt
\parskip=5pt plus 1pt minus 1pt
\tolerance 8000
\vfill\eject\immediate\closeout\reffile%\parindent=20pt
\centerline{{\bf References}}\bigskip\frenchspacing%
\input refs.tmp\vfill\eject\nonfrenchspacing

\bye